\documentclass[aps,amsmath,amssymb,reprint,superscriptaddress,a4paper,showpacs,showkeys,pre]{revtex4-1}
\usepackage{graphicx} 
\usepackage{dcolumn} 
\usepackage{bm}
\usepackage{hyperref} 
\usepackage{color}
\usepackage{amsmath}
\usepackage{amssymb}

\newcommand{\br}{\mathbf r}
\newcommand{\bq}{\mathbf q}

\begin{document}

\title{The Role of Contact Angle Hysteresis for Fluid Transport in Wet
  Granular Matter}

\author{Roman Mani} \affiliation{Computational Physics for Engineering
  Materials, ETH Zurich, 8093 Zurich, Switzerland}
\email[]{manir@ethz.ch}

\author{Ciro Semprebon} \affiliation{MPI for Dynamics and
  Self-Organization, G\"ottingen, Germany}

\author{Dirk Kadau} \affiliation{Computational Physics for Engineering
  Materials, ETH Zurich, 8093 Zurich, Switzerland}

\author{Hans J. Herrmann} \affiliation{Computational Physics for
  Engineering Materials, ETH Zurich, 8093 Zurich, Switzerland}
\affiliation{Departamento de F\'isica, Universidade Federal do
  Cear\'a, Fortaleza, Cear\'a 60451-970, Brazil}

\author{Martin Brinkmann} \affiliation{MPI for Dynamics and
  Self-Organization, G\"ottingen, Germany} \affiliation{Experimental
  Physics, Saarland University, Saarbr\"ucken, Germany}

\author{Stephan Herminghaus} \affiliation{MPI for Dynamics and
  Self-Organization, G\"ottingen, Germany}

\begin{abstract}  
  The stability of sand castles is determined by the structure of wet
  granulates. Experimental data about the size distribution of fluid
  pockets are ambiguous about their origin. We discovered that contact
  angle hysteresis plays a fundamental role in the equilibrium
  distribution of bridge volumes, and not geometrical disorder as
  commonly conjectured, which has substantial consequences on the
  mechanical properties of wet granular beds, including a history
  dependent rheology and lowered strength. Our findings are obtained
  using a novel model where the Laplace pressures, bridge volumes and
  contact angles are dynamical variables associated to the contact
  points. While accounting for contact line pinning, we track the
  temporal evolution of each bridge. We observe a cross-over to a
  power-law decay of the variance of capillary pressures at late times
  and a saturation of the variance of bridge volumes to a finite value
  connected to contact line pinning. Large scale simulations of liquid
  transport in the bridge network reveal that the equilibration
  dynamics at early times is well described by a mean field model. The
  spread of final bridge volumes can be directly related to the
  magnitude of contact angle hysteresis.
\end{abstract}

 \maketitle

\section{Introduction}
\label{sec:introduction}

Our daily life is strongly affected by the mechanical properties of
wet granulates, be it through the stability of the soil on which we
construct our buildings or the consistency of food we enjoy. The
structure of the fluid interfaces in a wet granular bed essentially
determines the strength of capillary cohesion and thus plays a key
role for the prediction and mitigation of devastating events such as
landslides or the clogging of industrial pipes
\cite{Herminghaus2013,Fernandez_Mani2013}. For a thorough
understanding, a realistic model of fluid movement is indispensable:
The capillary pressure driven fluid transport in wet granulates
\cite{Lukyanov2012,Scheel2008a} changes the fluid distribution which
in turn suggests that the mechanical properties are affected by the
fluid displacement.

Indeed, shear thinning for example is observed for partially saturated
beds of glass beads \cite{Herminghaus2005,Herminghaus2013,Kohonen2004}
due to a delayed reconfiguration of fluid interfaces. Knowledge about
the evolution of the fluid distribution in the granular assembly is
relevant to many industrial processes that involve coating of powders
and grains \cite{Teunou2002}, or mixing additives with particulate
materials \cite{Simons2000,Litster2004}. The precise microscopic
features which crucially determine the nature of the equilibration and
fluid transport processes are yet largely unexplored.

In this article we discover in simulations of the fluid equilibration
process in a bed of spherical beads that contact line pinning can have
a profound effect on the evolution of capillary bridge volumes. In
agreement with the majority experiments on fluid transport and
pressure equilibration in wet granular beds
\cite{Kohonen2004,Scheel2008a,Scheel2008b,Lukyanov2012}, we consider
the wetting fluid to be the minority phase forming capillary bridges
at grain contacts or in narrow gaps between neighboring grains. Our
simulations demonstrate that contact angle hysteresis broadens the
distribution of final bridge volumes while the polydispersity of bead
radii and the distribution of gap separations in the granular bed
appear to play only a minor role. The present approach accounts for
the discreteness of the bridge network and incorporates the history
dependence of the bridge shape caused by contact angle hysteresis.

Combining the model for fluid transport with the Contact Dynamics (CD)
algorithm \cite{Jean1992,Moreau1994,Brendel2004,Mani2012a} allows us
to simulate the motion of beads due to an external driving and the
redistribution of liquid after the external energy input has been
stopped. In this way, a direct comparison of our numerical results to
the evolution of bridge volumes using X-ray microtomography imaging of
fluid distribution measured by Scheel et
al.~\cite{Scheel2008a,Scheel2008b} is possible. In these experiments
vertical shaking of the granular bed was employed to distribute the
wetting liquid on the surface of the beads. Shortly after the
agitation of the wet granular bed (e.g.~by stirring of shaking) has
been stopped, the surfaces of the beads are covered with a `thick'
macroscopic film of the wetting fluid \cite{Kohonen2004}. Frequent
closing and opening of contacts leads to a uniform distribution on the
beads since the timescale of collisions is typically much smaller than
timescales of viscous flows on the surface of the beads or of
diffusion through the continuous fluid phase.

After the beads have settled into stable positions in the bead pack,
the wetting fluid starts of flow to the points of closest proximity of
opposing bead surfaces, forming macroscopic capillary bridges. At the
end of this process, the liquid volume varies from bridge to bridge
reflecting both the disorder in the bead pack and the distribution of
fluid patches on the beads immediately after shaking.

\begin{figure}[tb]
  \centering 
  \includegraphics[width=\columnwidth]{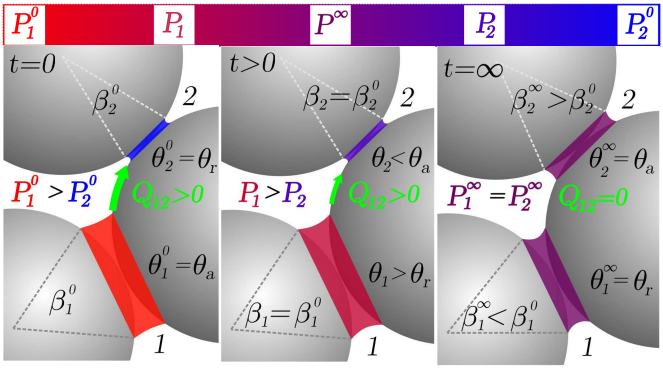}
  \caption{(Color online) Effect of contact angle hysteresis on the
    equilibration dynamics of two capillary bridges labeled $i=1,2$
    between beads with identical radii in mechanical contact.}
  \label{fig1}
\end{figure}

The initial regime of bridge formation is followed by an equilibration
of capillary pressure which must involve an exchange of wetting fluid
between neighboring bridges. In this `equilibration regime' the
majority of the wetting fluid will be contained in macroscopic
capillary bridges, cf.~the sketch in Fig.~\ref{fig1}. In the present
work we will focus on the collective evolution of bridge volumes in
the equilibration regime where we assume that volume conservation
holds for the wetting fluid. Our model for the equilibration dynamics
is based on the assumption that the fluid transport is entirely driven
by differences in the local capillary pressure. For wetting liquids
with a low vapor pressure this transport proceeds mainly through thin
wetting films in the roughness of the beads
\cite{Kohonen2004,Lukyanov2012}. A diffusive transport through the
vapor phase has to be considered for volatile liquids that exhibit a
sufficiently high vapor pressure \cite{Shahraeeni2010,
  Shahraeeni2012}.

Central aim of the present work is to quantify the impact of contact
angle hysteresis and polydispersity of the particles in the granular
bed onto the equilibration of bridge volumes. In particular, we
address the question for typical timescales of the transport
processes. These timescales of capillary pressure equilibration play
an important role in the rheology of partially saturated wet granular
beds.

\section{Physical model}
\label{sec:physical_model}

For simplicity, we consider the wetting liquid that forms the
capillary bridges (minority phase) to be surrounded by a continuous
vapor phase (majority phase). Analogous statements can be obtained for
binary mixtures of partially miscible liquids where the minority phase
is a wetting liquid while the majority phase is a non-wetting liquid.

Throughout the following considerations, we will neglect contributions
of the hydrostatic pressure to the pressure difference $P\equiv
p_l-p_v$ between the liquid (l) and vapor (v). Hence, the capillary
pressure $P$ is the same in every point of the liquid-vapor
interface. Provided a uniform interfacial tension $\gamma$, the
Young-Laplace equation
\begin{equation}
  P=2H\gamma~
  \label{eq:Laplace_equation}
\end{equation} 
implies that the interface is a surface of constant mean curvature
$H$. The mean curvature $H$ is the sum of the two principal curvatures
$\kappa_1$ and $\kappa_2$ in a respective point of the interface.

The influence of gravity or buoyancy on the capillary pressure $P$
will become apparent whenever the maximal difference in hydrostatic
pressure is of the order or exceeds $\gamma/R_0$, where $R_0$ is the
average bead radius. This is the case if the vertical extension of the
system becomes comparable, or exceeds the length
\begin{equation}
  L_\perp = \frac{\gamma}{\Delta\rho g\,R_0}~,
  \label{eq:vertical_extension}
\end{equation}
where $\Delta\rho\equiv \rho_l-\rho_v$ is the difference in mass
density between the wetting liquid $l$ and the ambient fluid phase
$v$, while $g$ is the acceleration of gravity.

\subsection{Liquid transport}
\label{subsec:liquid_transport}

Two different models for the local transport of wetting liquid between
neighboring capillary bridges can be considered. In case the wetting
liquids forms a thin film on the surface of the grains, any pair of
bridges on the same bead are hydraulic connected. Assuming a
quasi-steady flow in these films the total flux between two capillary
bridges on the same grain is proportional to the difference in their
capillary pressure. The corresponding liquid mobility, or conductance
coefficient depends on the geometry and distance of the three phase
contact lines of the bridges. Besides the advective transport
mechanism, we have to account for a capillary pressure equilibration
by mass diffusion of the wetting liquid through the ambient continuous
phase. The latter transport mechanism may be effective for wetting
liquids with a high vapor pressure or liquids with a partial
miscibility in the continuous liquid phase.

\subsubsection{Viscous film flow}
\label{subsubsec:viscous_flow}

For systems where the non-volatile liquid forms a wetting film on the
surface of the beads we will regard the average thickness $h_0$ of the
film to be insensitive to the pressure difference $P$ between the
wetting liquid and the vapor phase. This assumption is justified
whenever the typical horizontal length scales of the roughness is much
smaller than the radius of curvature of the menisci in the macroscopic
capillary bridges. In addition, we need to assume a sufficiently small
microscopic contact angle of the liquid-fluid interface on the rough
grain surfaces that permits the formation of a percolating liquid film
\cite{Herminghaus2012,Herminghaus2013}.

Assuming a stationary viscous flow in the thin film at any instance in
time, the volume flux $Q_{ij}$ between a pair of bridges $i$ and $j$
on the same beads will be proportional to the difference of capillary
pressures $P_i-P_j$ and the liquid mobility $\sim h_0^3/\eta$ in the
thin film \cite{DeGennes2004}:
\begin{equation}
  Q_{ij}=C_{ij}\frac{h_0^3}{\eta}\,(P_i-P_j)~,
  \label{eq:volume_flux_film}
\end{equation}
where $\eta$ is the dynamic viscosity of the wetting liquid, $h_0$ the
thickness of the wetting film, and $P_{i,j}$ the Laplace pressure of
the two neighboring bridges. Here, we assume that pressure gradients
occur over a typical distance of the order $\sim R_0$. Similarly, we
estimate the circumference of the three phase contact line of the
bridges on a bead to be of the order $\sim R_0$. The dependence on the
particular geometry of the neighboring capillary bridges is adsorbed
in the dimensionless conductance coefficient $C_{ij}$, being of the
order of unity.

An exact computation of the conductance coefficients requires
solutions to the Laplace equation of the local capillary pressure in
the film. Dirichlet boundary data for the capillary pressure are given
on the contact lines. Conformal mapping of the surface of the sphere
to a plane allows us to reduce the problem to the computation of the
capacitance per length of two parallel and infinite cylindrical
conductors \cite{Adams1962}. This results shows that the dependence on
the size and distance of the bridges is at most logarithmic.

\subsubsection{Diffusion through ambient fluid}
\label{subsubsec:diffusion}

A rough estimate of the diffusive flux of a volatile wetting liquid
through its ambient vapor phase can be derived from Fick's first law
\cite{Shahraeeni2010,Shahraeeni2012}. In full analogy to the second
regime of capillary equilibration in the thin film model described
above, we will assume that the concentration profile of liquid
molecules in the vapor phase has already leveled in every single pore
of the bead pack. Gradients in the concentration are noticeable only
on length scales much larger than the typical dimension of a pore. As
shown in the appendix \ref{sec:appendixA}, we can write Kelvin's
equation in the form
\begin{equation}
  n_v-n_v^\ast=\frac{P\,{n_v^\ast}^2}{p^\ast n_l^\ast}
  \label{eq:Kelvin}
\end{equation}
which allows us to express the particle density of liquid molecules
$n_v$ in the vapor phase through the vapor pressure $p^\ast$ and the
densities $n_v^\ast$ and $n_l^\ast$ at bulk coexistence, and the
pressure difference $P \equiv p_l-p_v$ across the curved interface. As
shown in the appendix \ref{sec:appendixA}, Kelvin's
equation~(\ref{eq:Kelvin}) still holds if we replace the vapor
pressure $p^\ast$ and densities $n_v^\ast$, $n_l^\ast$ at bulk
coexistence of the pure phases by the partial vapor pressure $\tilde
p_v$, and the particle densities $\tilde n_v$, $\tilde n_l$ for a
given atmospheric pressure $p_0$ of the gas phase, respectively.

Using Fick's first law, we can express the volume flux of liquid
caused by diffusion of liquid molecules between bridge $i$ and $j$ as
\begin{equation}
  Q_{ij}=\tilde C_{ij} \frac{ D\,{n_v^\ast}^2
    R_0}{p^\ast{n_l^\ast}^2~, } \left(P_i-P_j\right)~,
  \label{eq:volume_flux_vapor}
\end{equation}
where $D$ is the diffusion constant of molecules in the vapor phase,
cf.~\ref{sec:appendixA}. The dimensionless prefactor $\tilde C_{ij}$
accounts for the specific geometry of the pore space and the bridge
interfaces.
  
In full analogy to the dimensionless conductance coefficient $C_{ij}$
in the viscous film model, we expect the prefactor $\tilde C_{ij}$ to
be of order unity. A computation of the matrix elements $\tilde
C_{ij}$ involves solutions of the three dimensional Laplace equations
for the stationary density profiles with Dirichlet boundary data on
the liquid-vapor interface of the capillary bridges.

\subsubsection{Transport timescales}
\label{subsubsec:timescales}

In view of the numerical simulations, it is useful to
non-dimensionalize all relevant physical quantities. In the following
we employ the average bead radius $R_0$ as a unit of length. After
rescaling the pressure by $\gamma/R_0$, the relation
(\ref{eq:volume_flux_film}) for the volume flux yields a
capillary-viscous time scale of the form
\begin{equation}
  T_{\rm v} \equiv \frac{\eta R^4_0}{\gamma h_0^3}
  \label{eq:T0_viscous_flow}
\end{equation}
for the pressure equilibration by fluid transport through a thin
viscous film. Employing the relation (\ref{eq:volume_flux_vapor})
instead of (\ref{eq:volume_flux_film}) leads to a time scale
\begin{equation}
  T_{\rm d} \equiv \frac{ p^\ast {n_l^\ast}^2 R^3_0}{
    \gamma\,{n_v^\ast}^2 D}
  \label{eq:T0_diffusion}
\end{equation}
for a diffusive transport through the vapor phase.

Volume fluxes of both transport modes can be superimposed as long as
the linear relations between the respective volume flux and pressure
difference eqns.~(\ref{eq:volume_flux_film}) and
(\ref{eq:volume_flux_vapor}) are applicable. This implies that the
timescale
\begin{equation}
  T_0= \frac{T_{\rm v} T_{\rm d}}{T_{\rm v}+T_{\rm d}}
  \label{eq:T0_minimum}
\end{equation} 
accounts both for flows through thin liquid films and for diffusion
through the vapor phase. The corresponding dimensionless
characteristic number
\begin{equation}
  \mathbf{J}\equiv \frac{T_{\rm v}}{T_{\rm d}}= \frac{
    D\,\eta\,{n_v^\ast}^2 R_0}{ p^\ast\,{n_l^\ast}^2 h_0^3}
  \label{eq:J_number}
\end{equation}
allows us to discriminate between two different transport regimes: for
$\mathbf{J} \gg 1$, the diffusive transport through the vapor phase
dominates while for $\mathbf{J}\ll 1$ the main transport is by viscous
flows through the thin wetting film. It has been shown experimentally
that both cases may indeed be encountered
\cite{Musil1993,Herminghaus1995,Seemann2001}. The central message here
is that although the relative importance of the two mechanisms may
vary and can be expressed by the value of $\mathbf{J}$, the form of
the transport equations remains unchanged. Hence it is sufficient to
treat the film flow case, as we will do in what follows.

\subsection{Capillary bridges}
\label{subsec:capillary_bridges}

As in a previous Letter \cite{Mani2012a} the capillary pressure $P$ of
a bridge with volume $V$ spanning a gap with separation $S$ between
beads of equal radius $R_0$ is interpolated from tabulated values. To
account for contact angle hysteresis, we consider capillary bridges
with either pinned or freely moving contact lines. A suitable
parameter to describe the position of the pinned contact line of a
capillary bridge is the opening angle $\beta$, cf.~also the sketch in
Fig.~\ref{fig1}.

The capillary pressure of a bridge is a function $\tilde
P_0(V,S,\beta)$ for the bridges with pinned contact lines depending,
besides on the volume $V$ and gap separation $S$, on the opening angle
$\beta$. For bridges with a freely sliding contact line and a
prescribed contact angle $\theta$, the capillary pressure is expressed
as a function $P_0(V,S,\theta)$. In addition to the capillary
pressure, we compute the contact angle $\theta(V,S,\beta)$ of bridges
with a pinned contact line and the opening angle $\beta(V,S,\theta)$
for those with a sliding contact line. To numerically compute the
functions, $\tilde P_0(V,S,\beta)$, $P_0(V,S,\theta)$,
$\theta(V,S,\beta)$, and $\beta(V,S,\theta)$ we employed numerical
energy minimizations using the public domain software Surface Evolver
\cite{Brakke1996} where we considered only axially symmetric
interfacial profiles using an effectively two dimensional
representation of the bridge state. The rupture distance of a bridge
with fixed volume $V$ is computed from the approximate expression
\begin{equation}
  S_0^\ast=\left(1+\frac{\theta}{2}\right)V^{1/3}
  \label{eq:rupture_distance}
\end{equation}
as given by Willet et al.~\cite{Willet2000}. As shown in one of the
following works, Ref.~\cite{Willet2003},
eqn.~(\ref{eq:rupture_distance}) is still a good approximation for
bridges with a pinned contact line provided that the opening angle
$\beta$ is not too small. In this case, the actual contact angle
$\theta=\theta(V,S,\beta)$ of the bridge is used in
eqn.~(\ref{eq:rupture_distance}).
  
To account for the dependence of capillary pressure $P$ on the volume
$V$, the surface to surface separation $S$, and on the individual
radii $R_i$ of the two beads $i$=$1,2$ we use scaled quantities
\begin{eqnarray}
  P(V,S,\theta)&=&\xi^{-1}\,P_0(\xi^{-3}
  V,\xi^{-1}\,S,\theta)\\[0.8mm] \tilde
  P(V,S,\beta)&=&\xi^{-1}\,\tilde P_0(\xi^{-3}
  V,\xi^{-1}\,S,\beta)\\[0.8mm]
  S^\ast(V,\,\theta)&=&\xi\,S_0^\ast(\xi^{-3}\,V,\theta)
\end{eqnarray} 
where the dimensionless factor $\xi\equiv R_\mathrm{c}/R_0$ accounts
for the Derjaguin mean
\begin{equation}
  R_\mathrm{c}\equiv \frac{2R_1 R_2}{R_1+R_2}
\end{equation}
of bead radii \cite{Willet2000}.

\subsection{Contact angle hysteresis}
\label{subsec:hysteresis}

Roughness or variations in the chemical composition of the surface
lead to a history dependent contact angle \cite{DeGennes2004}.
Assuming ideal surfaces with uniform surface heterogeneities the
contact angle $\theta$ of a fluid interface in a mechanical
equilibrium falls into an interval $[\,\theta_{\rm r},\theta_{\rm a}]$
limited by the receding and advancing contact angle $\theta_{\rm r}$
and $\theta_{\rm a}$, respectively. The contact line starts to recede
once the local contact angle $\theta$ equals $\theta_{\rm
  r}$. Likewise, the contact line starts to advance if $\theta$ equals
$\theta_a$ For all intermediate values of $\theta \in [\,\theta_{\rm
    r}, \theta_{\rm a}]$ the contact line is immobilized,
i.e.~`pinned'. The magnitude of contact angle hysteresis is defined as
the width $\Delta \theta\equiv\theta_{\rm a}-\theta_{\rm r}$ of the
range of static contact angles.

The sketch in Fig.~\ref{fig1} illustrates the impact of contact angle
hysteresis on the equilibration dynamics for the example of two
capillary bridges $i=1,2$. Starting from an initial state with volumes
$V_1^0>V_2^0$, opening angles $\beta^0_1>\beta^0_2$ and contact angles
$\theta_1^0=\theta_{\rm a}$, $\theta_2^0=\theta_{\rm r}$ at $t=0$, the
capillary pressure difference $P_1^0-P_2^0>0 $ drives a fluid flux
$Q_{12}>0$ through the thin film between bridge $1$ and $2$. As the
total volume of bridge $1$ and $2$ must be conserved we have $\dot
V_1=-\dot V_2=-Q_{12}$, using the dot as a short hand notation for the
total time derivative. Although the bridge $1$ is shrinking and bridge
$2$ is growing, all contact lines remain first in a pinned state
($\beta_1=\beta_1^0$, $\beta_2=\beta_2^0$, and
$\theta_{1,2}\in[\,\theta_{\rm r}, \theta_{\rm a}]$). Only while
approaching the final state at $t\rightarrow \infty$ with
asymptotically equal pressures $P^\infty_1=P^\infty_2$, the contact
lines become depinned ($\theta_1^\infty=\theta_{\rm r}$,
$\theta_1^\infty=\theta_{\rm a}$), but still with bridge volumes
$V_1^\infty>V_2^\infty$. In the absence of contact hysteresis, one
would observe equal final volumes $V_1^\infty=V_2^\infty$.

\subsection{Network model}

The dynamics of fluid transport in the static bed of beads is studied
in a coarse grained network model where each capillary bridge forms a
node $i$ of the network. The nodes are connected by bonds $(i,j)$,
representing the thin film between bridges $j$ and $i$ on the same
bead. Mass conservation of the wetting fluid demands that the rate of
volume change of bridge $i$ is the sum
\begin{equation}
  \dot V_i = -\sum_{j\in {\cal N}(i)} Q_{ij} = \sum_{j\in {\cal N}(i)}
  C_{ij}\,(P_j- P_i)~,
  \label{eq:volume_evolution_network}
\end{equation}
over the set ${\cal N}(i)$ of all bridges $j$ that are connected to
bridge $i$. The relative geometry of a pair of neighboring bridges
$(i,j)$, the effect of other bridges, or individual bead radii is
collected in the dimensionless conductance coefficient $C_{ij}$, which
is of the order of one. The logarithmic dependence of $C_{ij}$ on the
minimum distance of contact lines applies only to capillary bridges
close to coalescence. Since we exclude this exceptional case, we set
$C_{ij}=1$ throughout.
  
For given derivatives $\dot V_i$ of bridge volumes
eqn.~(\ref{eq:volume_evolution_network}) and gap separations $\dot
S_i$, we are able to compute the total time derivatives of the opening
angle $\beta_i$ and contact angle $\theta_i$ of bridge $i$. Using the
differentials
\begin{eqnarray}
  a &\equiv& \partial_V \beta(V_i,S_i,\theta_i)\,\dot V_i+\partial_S
  \beta(V_i, S_i,\theta_i)\,\dot S_i \nonumber \\[0.5mm] b &\equiv&
  \partial_V \theta(V_i,S_i,\beta_i)\,\dot V_i+\partial_S \theta(V_i,
  S_i,\beta_i)\,\dot S_i \nonumber
\end{eqnarray}
we can express the time derivatives $\dot\theta_i$ and $\dot\beta_i$
as
\begin{eqnarray}
  \dot\beta_i=a~,\quad \dot\theta_i = 0 & \quad {\rm for} \quad&
  \left\{
  \begin{array}{ccc}
    a \geq 0 & {\rm and} & \theta_i=\theta_a \\[0.5mm]
    a \leq 0 & {\rm and} & \theta_i=\theta_r
  \end{array}	 
  \right.
  \label{eq:beta_dot} \\
  \dot\beta_i=0~,\quad \dot\theta_i = b & \quad {\rm for} \quad&
  \theta_r < \theta_i < \theta_a~.
  \label{eq:theta_dot}
\end{eqnarray}
The dependence on the gap separation $S$ in the differentials becomes
important if the beads are allowed to move relative to each other. In
a static packing, however, the angles $\beta_i$ and $\theta_i$ depend
on $V$, only.
  
According to eq.~(\ref{eq:beta_dot}), the contact line of a bridge $i$
can only slide outward ($\dot\beta_i>0$) for a contact angle
$\theta_i=\theta_a$. Capillary bridges with inward sliding contact
lines ($\dot\beta_i<0$), however, have to satisfy
$\theta_i=\theta_r$. Equations (\ref{eq:beta_dot}) guarantees that the
contact angle $\theta_i$ cannot become larger than $\theta_a$ or
smaller than $\theta_r$. The complementary case of an immobilized
contact line is taken into account by eqs.~(\ref{eq:theta_dot}): Any
intermediate value $\theta_r<\theta_i<\theta_a$ of the contact angle
$\theta_i$ implies a pinned contact line ($\dot\beta_i=0$).
  
As a consequence of the dynamics described by eqs.~(\ref{eq:beta_dot})
and (\ref{eq:theta_dot}), the state of a capillary bridge is not
uniquely determined by the volume $V$ and separation $S$. Depending on
whether the contact line is advancing, pinned, or receding we have to
employ different expressions for the capillary pressure:
\begin{equation}
  P_i = \left\{ 
  \begin{array}{ccc}
    P(V_i,S_i,\theta_a) & {\rm for} & \dot\beta_i>0 \\[0.8mm] 
    \tilde P(V_i,S_i,\beta_i) & {\rm for} & \dot\beta=0 \\[0.8mm] 
    P(V_i,S_i,\theta_r) & {\rm for} & \dot\beta_i<0
  \end{array}~.	
  \right.	   		
  \label{eq:P_dot}
  \end{equation}
The evolution of liquid volume $V_i(t)$, opening angle $\beta_i(t)$,
and contact angle $\theta_i(t)$ for each individual bridge $i$ is
obtained from an integration of
eqns.~(\ref{eq:volume_evolution_network}-\ref{eq:P_dot}).
  
\begin{figure}[tb]
  \begin{center}
    \includegraphics[width=0.95\columnwidth]{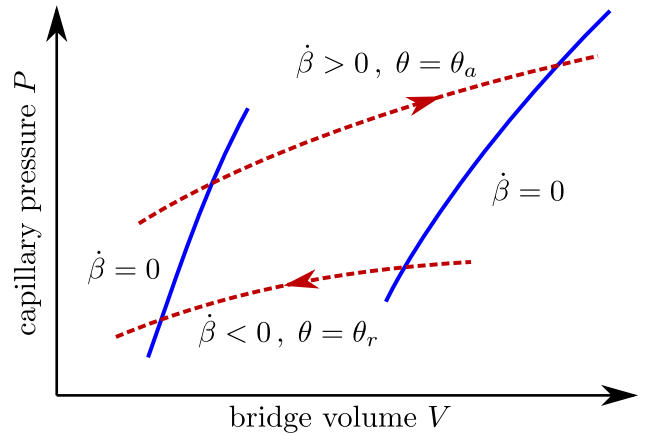}
    \caption{(Color online) Schematic of the capillary pressure $P$
      against the volume $V$ of a bridge. The lower and upper branch
      (dashed lines) corresponds to a receding (lower) and advancing
      (upper) contact line. The left and right branches (solid lines)
      describe bridges with a pinned contact line with small (left)
      and at large (right) opening angle.}
    \label{fig2}
  \end{center}
\end{figure}

The dependence of the capillary pressure $P$ on the history of bridge
volume $V$ for a fixed separation $S$ is illustrated in
Fig.~\ref{fig2}. The two flat branches in the sketch in
Fig.~\ref{fig2} correspond to the case of an advancing (solid line
top) and receding contact line (solid line bottom), while the step
curves (dashed lines) describe the pressure of a capillary bridge with
pinned contact lines. Owing to the history dependent dynamics
specified in eq.~(\ref{eq:beta_dot}), a capillary bridge moving on the
branch corresponding to an advancing contact line can be encountered
only for growing volumes. Similarly, bridges on the lower branch
corresponding to a receding contact line are encountered only for
shrinking volumes. In contrast, the volume of a bridge with a pinned
contact line can be either growing or shrinking.

A substantial simplification to the full network model
eqs.~(\ref{eq:volume_evolution_network}) can be obtained by replacing
the capillary pressure $P_j$ in
eqn.~(\ref{eq:volume_evolution_network}) with $j\in {\cal N}(i)$ by
the mean $\langle P \rangle$ taken over all bridges in the
network. The temporal evolution of individual bridge volumes $V_i$ in
this `mean field' approximation is
\begin{equation}
  \dot V_i=C\,\langle N_c \rangle (\langle P \rangle - P_i)
  \label{eq:mean_field_model}
\end{equation}
where $C$ is a global conductance coefficient and $\langle N_c
\rangle$ the average bridge coordination on a bead.

\subsection{Simulations}
\label{subsec:simulations}

In order to reproduce the experimental conditions of the equilibration
experiments in Refs.~\cite{Scheel2008a,Scheel2008b}, we simulate the
particle motion during the preparation step using the contact dynamics
(CD) algorithm. Since the beads employed in the experiments are a
sieving fraction of a wide size distribution it is reasonable to
assume bead radii which are uniformly distributed between a minimum
radius, $R_{-}$, and a maximum radius, $R_{+}$. The degree of
polydispersity is parameterized by the ratio $r_{pol} \equiv
R_{-}/R_{+}<1$, while $R_{+}$ is employed as our unit of length,
$R_0$. The open simulation box is bounded by a rigid movable wall at
the bottom $z=0$, applying periodic boundary conditions along the $x$
and $y$ directions. Gravity acts on the particles into $-z$ direction
with an acceleration of gravity $g=9.81$m/s$^2$.

To create a static packing similar to those in the time resolved x-ray
tomography experiments in Refs.~\cite{Scheel2008a,Scheel2008b}, the
bed of beads is fluidized by an oscillatory motion of the lower wall
of the container. The wetting fluid is redistributed exclusively via
rupture and formation of capillary bridges. If two beads touch each
other, a small bridge is created with $\theta=\theta_a$, where the
required, minimum fluid volume is provided by all neighboring
bridges. Similarly, if the bead separation exceeds the critical
distance $S^\ast(V)$, the bridge ruptures and the volume $V$ is
equally split and distributed onto all bridges on the two
beads. Capillary forces are small compared to the typical bead mass
and acceleration amplitudes during fluidization and are therefore
neglected.

\begin{figure}[tb]
  \begin{center}
    \includegraphics[width=\columnwidth]{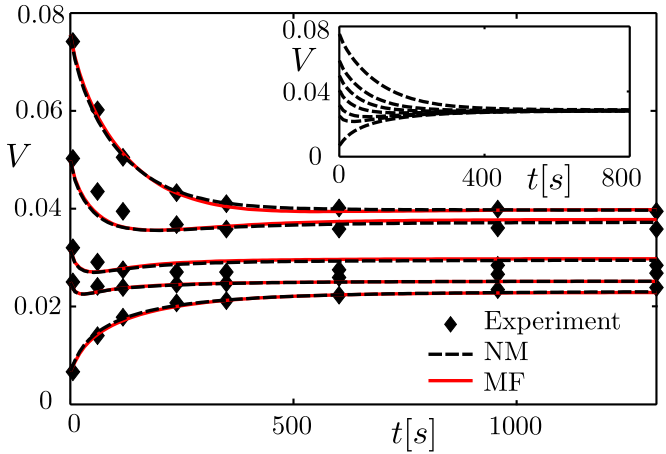}
    \caption{(Color online) Relaxation of individual capillary bridge
      volumes as a function of time in experiments (symbols) taken
      from Ref.~\cite{Scheel2008b} and in our simulations for $\Delta
      \theta=25^\circ$ in the full network model (`NW', black dashed
      lines) and mean field model (`MF', red solid lines). The inset
      shows the equilibration dynamics in the network model with
      $\Delta \theta=0^\circ$. The receding contact angle is $\theta_{\rm
        r}=7^\circ$ in both cases.}
    \label{fig3}
  \end{center}
\end{figure}

Once the agitation is stopped we let the beads settle by gravity into
a mechanically stable packing (setup A). The distribution of fluid
volumes obtained with this procedure is close to a decreasing
exponential function. To extend our calculations to larger systems
while saving the computational costs for the initial preparation, we
considered a three dimensional cubic lattice of beads in contact
applying periodic boundary conditions to the box with linear
dimensions $L=200$ (setup B). To mimic the initial state of bridges
after the preparation step in A, we initialized the bridges with an
exponentially decaying probability distribution function (PDF) of
volumes $\propto \exp(-V/\langle V \rangle)$ with a given average
volume $\langle V \rangle$ and set all contact angles
$\theta=\theta_a$. To follow the evolution of individual bridge
states, we integrated
eqs.~(\ref{eq:volume_evolution_network}-\ref{eq:P_dot}) for setup A
and B numerically in time using a simple forward Euler time scheme.

\section{Results and Discussion}
\label{sec:results}

The simulated evolution of bridge volumes in a static bed of beads
after shaking (set up A) is shown in Fig.~\ref{fig3}. Following the
analysis of the x-ray tomography data in Ref.~\cite{Scheel2008b}, we
classify each bridge according to their binned volume at time
$t=0$. Volume averages over all bridges being in the same bin at $t=0$
are plotted against the time $t$ elapsed after the agitation has been
stopped. To match the parameter of the experiments reported in
Ref.~\cite{Scheel2008b}, we set the polydispersity to $r_{pol}=0.8$
and chose a total fluid volume corresponding to an overall fluid
content of $W=2\times 10^{-2}$ of the wetting fluid with respect to
the total sample volume. The receding contact angle in our simulations
is set to $\theta_r=7^\circ$ throughout, being the smallest value that
allows for a reliable interpolation between the tabled functions
$P(S,\beta,\theta)$ and $V(S,\beta,\theta)$. The distribution of final
bridges volumes for an advancing contact angle $\theta_a=32^\circ$
(corresponding to a contact angle hysteresis of
$\Delta\theta=25^\circ$) provides the best match to the experimental
data in Ref.~\cite{Scheel2008b}.

The plot in Fig.~\ref{fig3} demonstrate that the average bridge volume
of a bin does not need to be a monotonously increasing or decreasing
function of time. In particular, we observe that some capillary
bridges with intermediate volume first shrink for a short time and
then tend to grow again. The inset of Fig.~\ref{fig3} shows the case
of a vanishing contact angle hysteresis $\Delta\theta=0^\circ$,
i.e.~for contact angles $\theta_a=\theta_r=7^\circ$. For
asymptotically large times all bridge volumes $V_i$ virtually converge
to the same value $\langle V \rangle=V^\infty$. This indicates that
local the geometry around contact points and small gaps, i.e.~the
possible locations of capillary bridges, are very similar. Because the
capillary pressure will finally be identical in all bridges, the final
value $V^\infty$ is determined only by the contact angle, the number
of contacts, and the total liquid volume in the granular bed.

\subsection{Transport timescales}
\label{eq:transport_timescales}

In addition to the values the contact angles $\theta_a$ and
$\theta_r$, we can employ the time scale $T_0$ as a free parameter to
fit the results of our numerical simulations to the experimental
data. The best fit between the data of the equilibration experiments
in Ref.~\cite{Scheel2008b} and the simulations in set up A shown in
Fig.~\ref{fig3} is obtained for $T_0=2.1\times 10^5$s. Rather
unexpected, the volume equilibration in set up A appears to be almost
completed already after $t \gtrsim 10^{-3}$ expressed in the
dimensionless units of our simulations. Rescaling this time by $T_0$
leads to $\approx 250\,$s for the cross-over which is corroborated by
visual inspection of Fig.~\ref{fig3}.

If we assume that the transport proceeds exclusively through viscous
flows in thin films, we can employ eqn.~(\ref{eq:T0_viscous_flow}) to
estimate the equivalent thickness of the thin film. Setting the bead
radius to $R_0=280\,\mu$m, the interfacial tension and dynamic
viscosity of the aqueous ZnI$_2$ solution to $\gamma=72$\,mN\,m$^{-1}$
and $\eta=10^{-3}$Pa s, respectively, we obtain an estimate of
$h_0=74\,$nm for the effective thickness of the thin wetting
film. This values agrees well with the measured RMS roughness of glass
beads by Utermann et al.~\cite{Utermann2011} and the value of $h_0$
proposed by Lukyanov et al.~for their system of the non-volatile
liquid Tris(2-ethylhexyl) phosphate in a pack or quartz grains (Ottawa
sand)~\cite{Lukyanov2012}.

An estimate of the equilibration timescale $T_0$ in presence of a
purely diffusive transport in the vapor phase can be obtained from
eqn.~(\ref{eq:T0_diffusion}). Assuming pure water as a wetting liquid
and ambient conditions with a temperature of $300\,$K, we set the
vapor pressure to $p^\star=3.52\times 10^3\,$Pa, the molar densities
to $n_l^\star=5.56\times 10^{4}$mol m$^{-3}$ and $n_v^\star=1.41\,$mol
m$^{-3}$, the diffusion constant to $D=2.57\times
10^{-5}\,$m$^2$s$^{-1}$, and obtain a timescale of $T_d=6.5\times
10^7\,$s. The low value of the characteristic number $J=3.23\times
10^{-3}$ according to eqn.~(\ref{eq:J_number}) suggests that a
transport by thin film flow is favored over a diffusive transport in
the vapor phase. A further quantitative discussion requires the
numerical computation of the matrix elements $C_{ij}$ and $\tilde
C_{ij}$, and is left to future work.

A further argument that supports viscous flows as the dominant
transport mode in Ref.~\cite{Scheel2008b} is due to the high molarity
of the aqueous ZnI$_2$ solution. Any diffusive exchange of only the
volatile solvent between bridges, excluding and exchange of the
non-volatile ionic solute leads to gradients in chemical potential of
the solvent molecules. A simple estimate using Raoult's law
demonstrates that any gradient in the chemical potential due to
differences in capillary pressure is small compared to the gradients
caused by differences in molarities of the solute. Hence, a diffusive
exchange of bridge volumes is largely suppressed for high salt
concentrations.

\begin{figure}
  \centering \includegraphics[width=\columnwidth]{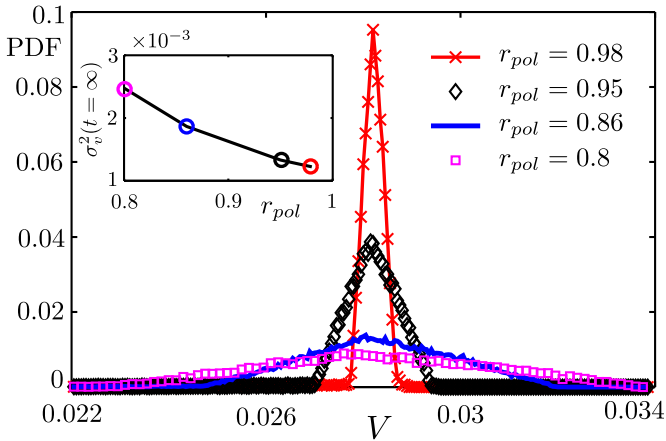}
  \caption{(Color online) Probability distribution function (PDF) of
    final bridge volumes $V$ for vanishing contact angle hysteresis
    according to set up A. Contact angle $\theta=7^\circ$. Inset:
    Variance of volumes in the final state for different
    polydispersities $r_{pol}$.}
  \label{fig4}
\end{figure}

\subsection{Final volume distribution}
\label{subsec:final_volume}

An explanation of the width of final bridge volumes by hydrostatic
pressure can be excluded for the equilibration experiments by Scheel
et al.~\cite{Scheel2008b}. Given the interfacial tension $\gamma$ and
density difference $\Delta \rho$ of the $1$M aqueous ZnI$_2$ solution
to the surrounding air, we compute from
eqn.~(\ref{eq:vertical_extension}) a maximal vertical extension of
$L_\perp \approx 1$\,cm. Since the analyzed field of view in the x-ray
tomography imaging had a height of less than $\ell\approx 5$mm
\cite{Scheel2008b}, we conclude that only differences in the Laplace
pressure were driving the exchange of wetting fluid between
neighboring bridges. Gradients of the final distribution of bridge
volumes caused by a mismatch in the hydrostatic pressure in the bulk
phases will contribute to the width of the volume distribution only if
$\ell \gtrsim L_\perp$ holds.

A remaining finite width of the final volume distribution can be
related to the degree of polydispersity $r_{pol}$ of bead
radii. Figure \ref{fig4} shows the probability density functions of
bridge volumes $V$ obtained in our simulations for a number of size
polydispersities $r_{pol}$ at late times. Apparently, all PDFs
approach a characteristic triangular shape while their width is
increasing as $r_{pol}$ becomes smaller. As expected, the variance of
bridge volumes, $\sigma_v \equiv (\langle V^2\rangle-\langle V
\rangle^2)^{1/2}$ that quantify the width of the PDFs, increases
monotonously with $r_{pol}$, the degree of polydispersity (cf.~the
inset of Fig.~\ref{fig3}). This weak sensitivity is a clear indication
that the final distributions of volumes observed in the experiments
cannot be explained by the polydispersity of bead radii
alone. Capillary bridges between spheres not in mechanical contact are
rare in packings prepared by setup A and the PDF of gap separations
$S$ between neighboring beads has a negligible impact on the final
volume distribution.

\begin{figure}
  \centering
  \includegraphics[width=\columnwidth]{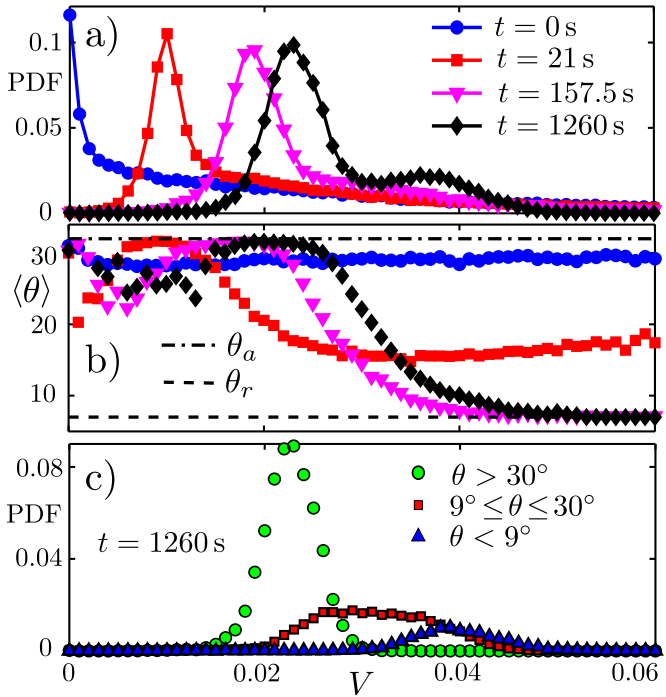}
  \caption{(Color online) Distributions for data in network model (NW)
    shown in Fig.~\ref{fig3}: a) PDF of bridge volume $V$ and b)
    Average contact angle $\langle \theta \rangle$ of bridges with
    volume $V$. c) PDF to find bridges in bins of contact angles
    $\theta$ close to the advancing, receding, or intermediate range
    at different times $t$.}
  \label{fig5}
\end{figure}

An obvious source for the finite width of the final distribution of
bridge volumes is a history dependent contact angle.
Figure~\ref{fig5}a) shows the PDF of bridge volumes at different times
after the agitation has been stopped for a finite contact angle
hysteresis of $\Delta\theta=25^\circ$. The raising of a second peak in
the PDF at large values from an initially monotonously decaying
function can be explained by a growing population of bridges with
receding contact lines. Shortly after preparation, all bridges exhibit
a contact angle close to $\theta_{\rm a}$, and the average contact
angle $\langle \theta \rangle$ of bridges with large volume gradually
decreases with time, cf.~Fig~\ref{fig5}~b). Bridges with large volume
and a capillary pressure close to zero shrink at a smaller rate
compared to the rate at which small bridges with advancing contact
angles and large negative pressure grow.  In the final state, the
majority of bridges is still in the advancing state and make up for
the large peak toward smaller volumes, as can be seen from
Fig.~\ref{fig5}~c). Most of the remaining bridges exhibit pinned or
receding contact lines, accounting for the second, smaller peak toward
larger volumes.

\subsection{Large time asymptotics}
\label{subsec:large_time_asymptotics}

Inspection of the direct comparison of the full network model, the
mean field model and the experimental data of Ref.~\cite{Scheel2008b}
in Fig.~\ref{fig3} demonstrates that the mean field model predicts the
dynamics for the chosen contact angle hysteresis of
$\Delta\theta=25^\circ$ at early times rather well. In order to
further assess the range of validity of the mean field model, and for
a quantities comparison of the evolution of capillary pressures and
bridge volume at late times we will from now on consider the larger
setup B.

\begin{figure}
  \centering 
  \includegraphics[width=\columnwidth]{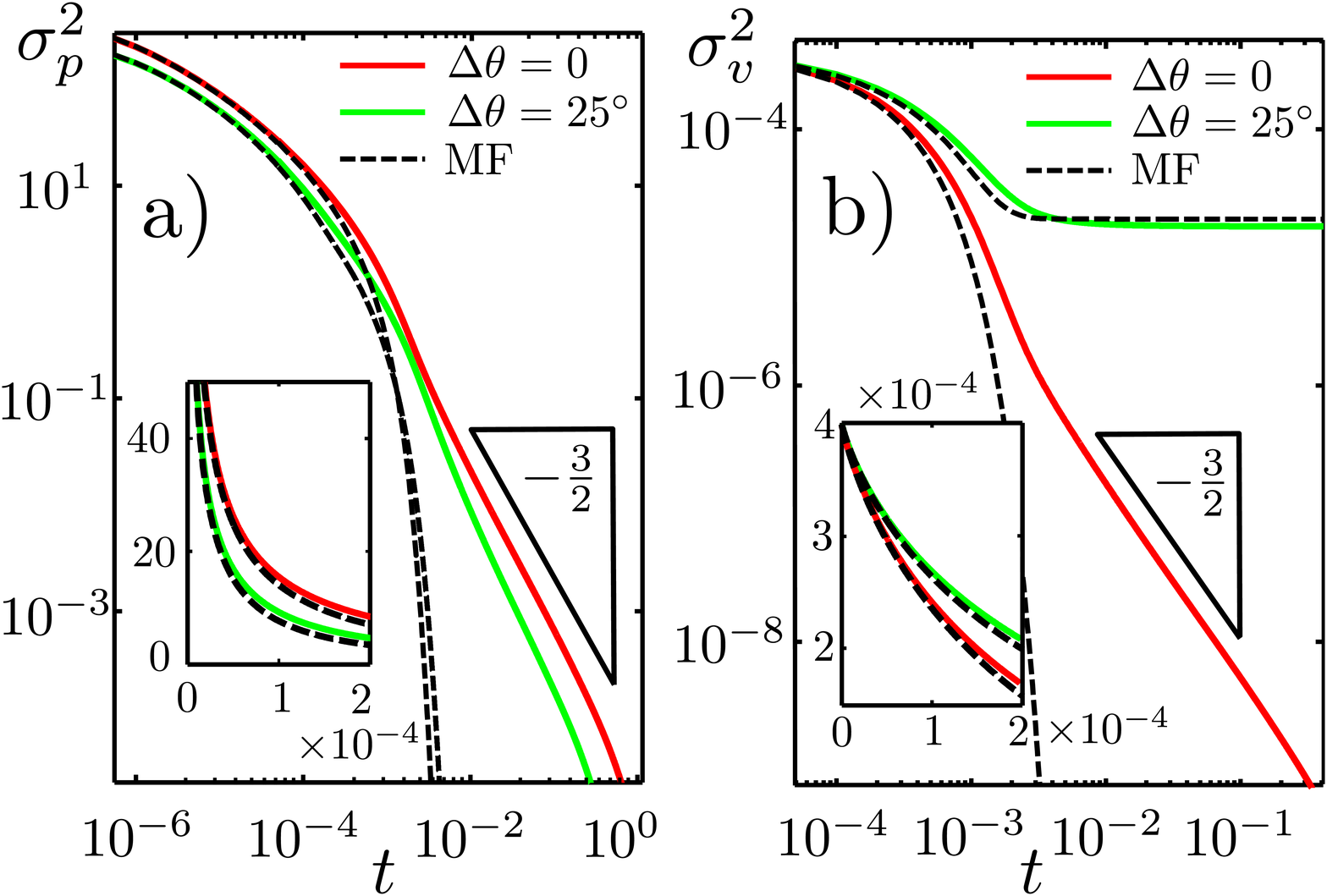}
  \caption{(Color online) a) Squared variance of capillary pressure,
    $\sigma_p^2$, and b) of bridge volumes, $\sigma_v^2$, as a
    function of time $t$ for $\Delta\theta=0^\circ$ (red solid lines) and
    $\Delta\theta=25^\circ$ in the network model and in the mean field
    approximation (dashed lines) for setup B. Insets show same data at
    early times in linear scale.}
  \label{fig6}
\end{figure}

Figure \ref{fig6}~a) and \ref{fig6}~b) display a comparison of the
square of variances $\sigma_p$ and $\sigma_v$ corresponding to the PDF
of capillary pressure $P$ and bridge volumes $V$, respectively, in the
mean field model (dashed lines) and the full network model (solid
lines). In contrast to the data shown in Fig.~\ref{fig3}, we use the
non-dimensionalized time in Fig.~\ref{fig6}. As for the simulations of
set up A we chose a contact angle hysteresis of $\Delta\theta=0^\circ$
(red solid lines) and $25^\circ$ (green solid lines).

In case of a vanishing contact angle hysteresis, we find a power-law
decay of both $\sigma_p^2$ and $\sigma_v^2$ in $t$ with an exponent
$-3/2$, as expected for continuum models for linear diffusion in three
spatial dimensions, cf.~the red solid lines in Fig.~\ref{fig6}~a) and
\ref{fig6}~b) \cite{Skhiri2012}. The cross-over from the initial
exponential decay to the power-law decay can be observed at an
estimated time $t=\tau_{\rm e}\approx 10^{-3}$. This is good agreement
with the dimensionless cross-over time observed in experiments and in
set up A. At early times $t\ll \tau_{\rm e}$ the evolution of both
$\sigma_p^2$ and $\sigma_v^2$ observed in the full network model is
well captured by the mean field model. This is not surprising as the
capillary pressure of bridges is not correlated immediately after the
initialization. The build-up of spatial correlations in the capillary
pressure at later times $t \gg \tau_{\rm e}$ will invalidate the mean
field model.

The decay of the squared variance $\sigma_p^2$ for a contact angle
hysteresis $\Delta\theta=25^\circ$ is slightly faster than for the
case of vanishing contact angle hysteresis (cf. the green and red
solid line in Fig.~\ref{fig6}~a). A pronounced qualitative difference,
however is observed for the squared variance of volumes $\sigma_v^2$,
displayed as the green and red solid lines in Fig.~\ref{fig6}~b).

Inspection of Fig.~\ref{fig6}~b) shows that the volume fluctuations in
the initial bridge volume do not fully decay in the limit
$t\rightarrow \infty$ if the contact angle hysteresis is set to the
finite value $\Delta\theta=25^\circ$. Instead the variance of bridge
volumes shown by the green solid curve in Fig.~\ref{fig6}~b) saturates
to a value $\sigma_v^2(t=\infty)>0$. Apparently, the time at the
cross-over and the level of the plateau is reasonably well captured by
the mean field model (dashed black line). Here, we would expect to
observe a break-down of the mean field model at late times because the
capillary pressure fluctuations become correlated irrespective of the
magnitude of contact angle hysteresis. The history dependence of the
contact angle, however, prevents a build-up of large correlations in
the fluctuations of neighboring bridge volumes. The time of the
cross-over of $\sigma_v^2$ to the plateau is comparable to $\tau_{\rm
  e}$ obtained from simulations for $\Delta \theta=0^\circ$.

\subsection{Cross over time}
\label{subsec:cross_over_time}

The cross-over time $\tau_e$ for the case without contact angle
hysteresis can be related to the time when the capillary pressure of
neighboring bridges starts to be correlated. To obtain an estimate for
$\tau_e$ we employ the continuum picture of a mode expansion of volume
fluctuations into a spectrum of modes with short and long
wavelengths. Similar arguments have been applied to determine the
cross-over time of diffusive solvent transport in a network of densely
packed emulsion droplets \cite{Skhiri2012}. Provided the equilibration
dynamics is described by a linear equation, we can think of the
spatial fluctuations of bridge volumes as a superposition of modes in
the local liquid saturation $c$. According to the Nyquist-Shannon
sampling theorem \cite{Goodman1962}, the shortest meaningful
wavelength of such a mode $c(\br,t)$ is given by twice the smallest
distance of two neighboring bridges. Toward large wavelengths, the
spectrum is limited by the system size $L$.

Solutions of the linear diffusion equation $D_e \Delta c = \partial_t
c$, with the three dimensional Laplacian $\Delta$ and the effective
diffusion constant $D_e$, show that amplitudes $c_\bq$ of a mode
$c(\br,t)=c_\bq\exp(i\,\bq\cdot\br- t/\tau_\bq)$ decay exponentially
with a timescale $\tau_\bq = D_e^{-1}\,|\bq|^{-2}$. The characteristic
decay time of the mode with the largest wave number $q_\mathrm{max}$
provides us with an estimate of $\tau_e$. Employing $2R_0$ as the
typical distance between two neighboring bridges we have
$q_\mathrm{max} \approx \pi/4R_0$.

An estimate of the effective diffusion constant $D_e$ can be obtained
from the link between the continuum Diffusion equation and the
discrete network model
eqn.~(\ref{eq:volume_evolution_network}). Considering the right hand
side of eqn.~(\ref{eq:volume_evolution_network}) as an approximation
of the Laplace operator we need to correct the left hand side of
(\ref{eq:volume_evolution_network}) for a numerical prefactor given by
the coordination of neighboring bridges for a proper mapping onto the
continuum diffusion equation. For the regular bridge network of setup
B we find a number of $N_c=10$ neighbors. When expanding the capillary
pressure $P_i$ in the volume fluctuation around the final volume
$V^\infty$ up to linear order we obtain a linear diffusion equation
for the fluctuations of the saturation $c(\br,t)$.

After collecting all factors, we find that the effective diffusion
constant $D_e$ of volume fluctuation in the bridge network is given by
an expression
\begin{equation}
  D_e \approx C\,\frac{N_c}{2d}
  \left.\frac{\partial P}{\partial V}\right|_{V=V^\infty}~,
  \label{eq:diffusion_constant}
\end{equation} 
where $d$ is the spatial dimension of the bridge network. Setting
global parameters separation $S=0$, contact angle $\theta=7^\circ$,
asymptotic volume $V^\infty \approx 0.02$ and the conductance
coefficient $C=1$, we obtain an estimated effective diffusion constant
$D_e\approx 500$ and a corresponding timescale $\tau_e\approx 10^{-3}$
which compares well with the observed cross-over time in simulation of
set-up~B, cf.~the beginning of the
section~\ref{eq:transport_timescales}. We find a faster diffusive
spreading in the network with a diffusion constant $D_e \approx 10^3$
if we consider capillary bridges with a pinned contact line and
opening angle $\beta=21.5^\circ$ (corresponding to a volume of
$V^\infty = 0.02$). We obtain this estimate by replacing the capillary
pressure $P$ in eqn.~(\ref{eq:diffusion_constant}) for bridges with
freely sliding contact lines at a fixed contact angle with the
pressure $\tilde P$ for bridges with pinned contact lines. An enlarged
value of the derivatives of the capillary pressure with respect the
volume is also apparent in the sketch in Fig.~\ref{fig2}. The
accelerated diffusive transport of liquid between capillary bridges
with pinned contact lines becomes even more pronounced for larger
contact angles.

\begin{figure}
  \centering \includegraphics[width=\columnwidth]{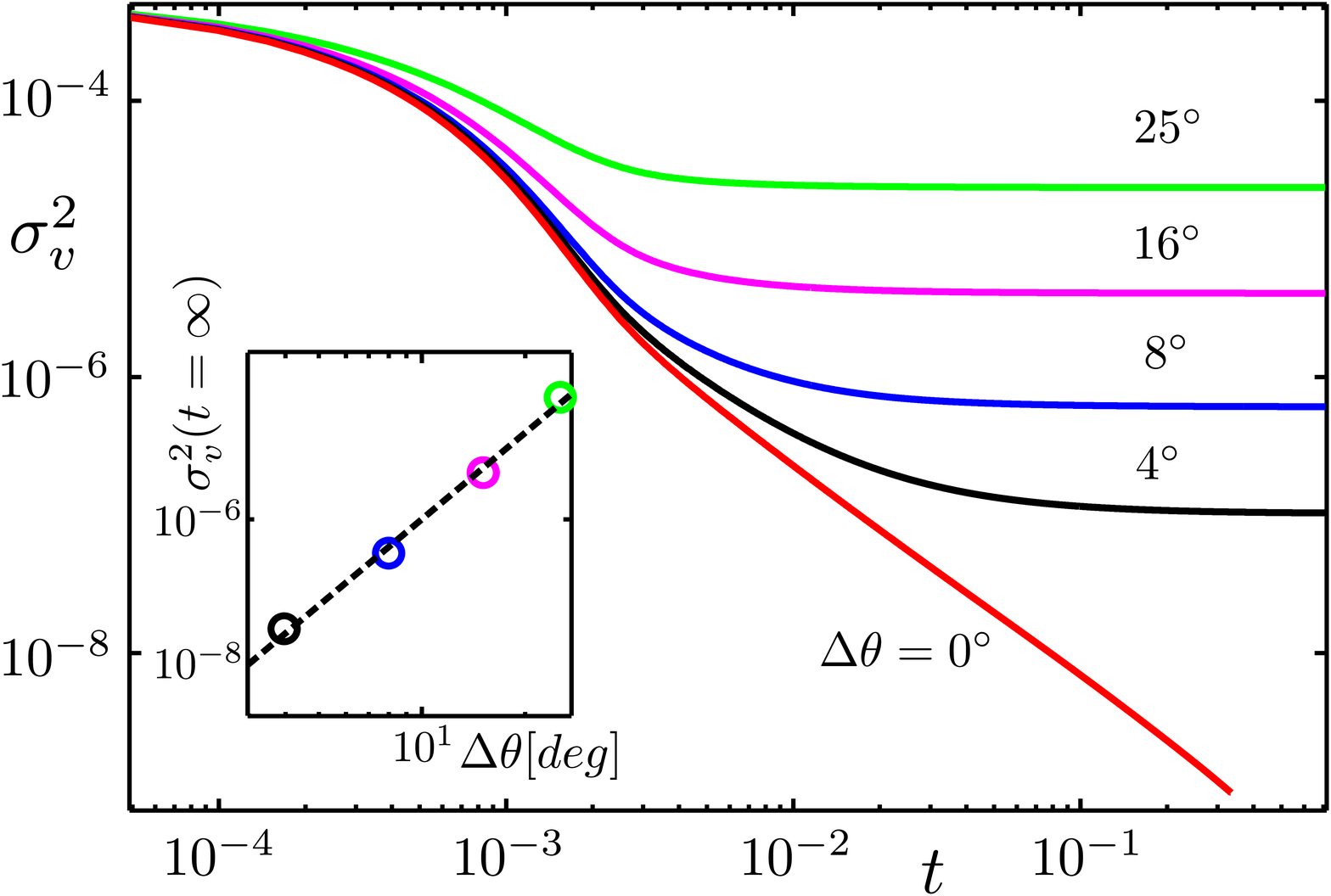}
  \caption{(Color online) Temporal evolution of the squared variance
    $\sigma_v^2$ of bridge volumes in the full network model for
    different magnitude of contact angle hysteresis
    $\Delta\theta=0^\circ, 4^\circ, 8^\circ, 15^\circ,
    25^\circ$. Inset: Double logarithmic plot of
    $\sigma_v^2(t=\infty)$ against $\Delta\theta$ in comparison to a
    power law $\sim \Delta\theta^3$ (dashed line).}
  \label{fig7}
\end{figure}

So far, we have explored the differences in the late time regimes of
equilibration of bridge volume for the two extreme cases of a large
contact angle hysteresis $\Delta \theta=25^\circ$ and vanishing
hysteresis. A cross-over to a power-law scaling of the variance of
bridge volumes observed for the ideal case $\Delta \theta=0^\circ$ can
be understood in the framework of a continuum model for diffusion. To
quantify the intermediate cases we considered the late time regimes
for a series of contact angle hysteresis. Figure \ref{fig7} shows the
decay of the square of the volume variance, $\sigma_v^2$, for
additional values $\Delta\theta=4^\circ,\,8^\circ,\,15^\circ$ in set
up B. The plot in the inset of Fig.~\ref{fig7} displays the value
$\sigma_v^2(t=\infty)$ at saturation in a double logarithmic plot. The
available data suggest a power law of the $\sigma_v^2(t=\infty)
\propto \Delta\theta^3$ that needs to be confirmed in further
investigations including larger regions of the parameter space.

\section{Outlook and conclusions}
\label{sec:conclusions}

The present numerical study demonstrates that contact angle hysteresis
strongly affects the evolution and distribution of capillary bridge
volumes after mixing a wetting fluid into a bed of spherical
beads. Our model study shows that disorder of the local geometry in
the granular bed such as the distribution of gap separations and the
polydispersity of bead radii hardly contribute to the final width of
bridge volume distribution. Unless the difference between the
advancing and receding contact angle on the beads is unrealistically
small, contact angle hysteresis will always dominate the final
distribution of bridge volumes.

Our numerical results further demonstrate that a mean field model for
the capillary pressure is sufficient to describe the equilibration
process at early times. A reliable prediction of the final
distribution of bridge volumes, however, must account for the network
of communicating bridges. The time of the cross-over between the early
time regime with negligible spatial correlations of the volume
fluctuations to a late time regime where a continuum model
describes the volume distribution can be derived for the ideal
case of a vanishing contact angle hysteresis. This cross-over time
provides an estimate for the saturation time of spatial volume
fluctuations in case of high contact angle hysteresis.

The network model can be further employed to address the role of
contact angle hysteresis in fluid transport in slowly sheared granular
beds, or the dynamics of liquid equilibration in particulate materials
driven by gradients in the local saturation. It can be easily extended
to account for flows of the wetting fluid induced by gravity or
evaporation.

\section{appendix}
\label{sec:appendixA}

\begin{appendix}
  
  Starting point for the derivation of eqns.~(\ref{eq:Kelvin}) and
  (\ref{eq:volume_flux_vapor}) for a one component system is the
  condition of coexistence
  \begin{equation}
    \mu_v(p_v,T)=\mu_l(p_l,T)
    \label{eq:chemical_potential_1}
  \end{equation}
  for the chemical potential $\mu_i$ of bulk phases $i\in\{v,l\}$ at
  temperature $T$. Due to the curved interface between the liquid and
  vapor phase, we have $p_l\neq p_v$. Expanding the chemical potential
  $\mu_i(p,T)$ of the liquid and vapor phase around the pressure
  $p^\star$ at coexistence with a flat interface leads to
  \begin{equation} \Delta \mu_i(p_i,T)\approx \mu^\ast(T)
    +\left.\frac{\partial \mu_i}{\partial
      p}\right|_{p=p^\ast}\!\!\!\Delta p_i~.
    \label{eq:expansion_mu}
  \end{equation}
with $\Delta p_i=p_i-p^\ast$.
  
Maxwell's relation for the grand potential
$G_i(p_i,N,T)=N_i\mu_i(p_i,T)$ of the bulk phases $i\in\{v,l\}$
provides us with the expression
\begin{equation}
  \frac{\partial \mu_i}{\partial p}=\frac{1}{n_i}
\end{equation}
for the molar density $n_i$ of the bulk phases, and we can rewrite
eqn.~(\ref{eq:expansion_mu}) in the form
\begin{equation}
  p_l-p_v=\frac{n_l^\ast-n_v^\ast}{n_v^\ast}\;\Delta p_v\approx
  \frac{n_l^\ast}{n_v^\ast}\;\Delta p_v~,
  \label{eq:pressure_diff_app}
  \end{equation}
where the last approximation is justified as long as
$n_l^\ast/n_v^\ast \gg 1$ holds. Since we assume the dilute vapor to
be an ideal gas, we can employ the relation $p_v/p^\ast=n_v/n_v^\ast$
and rewrite eqn.~(\ref{eq:pressure_diff_app}) in the form
\begin{equation}
    n_v-n_v^\ast=\frac{P\,{n_v^\ast}^2}{p^\ast\,n_l^\ast}~,
    \label{eq:Kelvin_app_1}
  \end{equation}
  in terms of the capillary pressure $P \equiv p_l-p_v$. Equation
  (\ref{eq:Kelvin_app_1}) is equivalent to Kelvin's equation.
  
  In most cases, the liquid coexists with a mixture of the liquid
  vapor and an inert gas that is insoluble in the liquid phase. In
  this case we need to replace the equilibrium condition
  eqn.~(\ref{eq:chemical_potential_1}) by
  \begin{equation}
    \mu_v(p_v,T)=\mu_l(p_0+P,T)
    \label{eq:chemical_potential_2}
  \end{equation}
  where $p_0$ is the ambient atmospheric pressure, $\tilde p_v$ the
  partial pressure of the liquid vapor. An expansion of both sides of
  eqn.~(\ref{eq:chemical_potential_2}) around the partial vapor
  pressure $\tilde p_v$ of a flat interface defined by
  \begin{equation}
    \mu_v(\tilde p_v,T)=\mu_l(p_0,T)
    \label{eq:definition_vapor_pressure}
  \end{equation}
  provides us with the Kelvin equation
  \begin{equation}
    n_v-\tilde n_v =\frac{P\,\tilde n_v^2}{\tilde p_v \tilde n_l}~,
    \label{eq:Kelvin_app_2}
  \end{equation}
  which is of the same form as eqn.~(\ref{eq:Kelvin_app_1}). All bulk
  quantity occuring in eqn.~(\ref{eq:Kelvin_app_2}), however, refer to
  a constant ambient pressure $p_0$ of the gas phase.
  
  According to Fick's first law of diffusion, the current $\mathbf{j}$
  of liquid molecules in the vapor phase can be related to the
  concentration gradient $\boldsymbol{\nabla}n_v$ and the diffusion
  constant $D$ by
  \begin{equation} 
    \mathbf{j}=-D\,\boldsymbol{\nabla}\,n_v~.
  \end{equation}
  Since the total particle current $\dot N_{ij}$ between bridge $i$
  and $j$ scales with the interfacial area of the bridges $\sim R_0^2$
  and with the gradient $|\boldsymbol{\nabla}|\sim R_0^{-1}$, we
  finally arrive a at volume flux of the liquid phase
  \begin{equation} 
    Q_{ij}= \tilde C_{ij}
    \frac{D\,{n_v^\ast}^2\,R_0}{p^\ast\,{n_l^\ast}^2}\,(P_i-P_j)~,
  \end{equation}
  where $R_0$ is the bead radius and $\tilde C_{ij}$ a numerical
  prefactor that accounts for the particular geometry of the bridges
  and pore space.
  
\end{appendix}

\begin{acknowledgments}
  We thank Mario Scheel and Jean-Christophe Baret for fruitful
  discussions and we acknowledge funding by the German Research
  Foundation (DFG) through the grants no.~He 2732/11-3 and He
  2016/14-3 in the SPP 1486 `PiKo' for funding. This work was also
  supported by grant number FP7-319968 Flow-CCS of the European
  Research Council (ERC) advanced grant.
\end{acknowledgments}

\end{document}